\documentclass{iau} 
\usepackage{graphicx}
\usepackage{hyperref}

\title[] 
{Study of Interplanetary and Geomagnetic Response of Filament Associated CMEs}

\author[Kunjal Dave \& Wageesh Mishra \& Nandita Srivastava \& R. M. Jadhav]   
{Kunjal Dave$^1$, Wageesh Mishra$^2$, Nandita Srivastava$^3$
 \and R. M. Jadhav$^4$}

\affiliation{$^1$C. U. Shah University, Wadhwan, Surendranagar-363030, Gujarat, India \\ email: {\tt kunjaldave88@gmail.com} \\[\affilskip]
$^2$University of Science and Technology of China, Hefei, Anhui-230026, China \\[\affilskip]
$^3$Udaipur Solar Observatory, Physical Research Laboratory, Udaipur-313001, India \\ email: {\tt nandita@prl.res.in} \\[\affilskip] 
$^4$Gujarat Arts and Science College, Ahmedabad- 380006, Gujarat, India }

\pubyear{2018}
\volume{340}  
\jname{Long-Term Datasets for the Understanding of Solar \\ and Stellar Magnetic Cycle}
\editors{Dipankar Banerjee, Jie Jiang, Kanya Kusano, \& Sami Solanki}
\begin{document}

\maketitle

\begin{abstract}
It has been established that Coronal Mass Ejections (CMEs) may have significant impact on terrestrial magnetic field and lead to space weather events. In the present study, we selected several CMEs which are associated with filament eruptions on the Sun. We attempt to identify the presence of filament material within ICME at 1AU. We discuss how different ICMEs associated with filaments lead to moderate or major geomagnetic activity on their arrival at the Earth. Our study also highlights the difficulties in identifying the filament material at 1AU within isolated and in interacting CMEs.
\end{abstract}

\firstsection 
\section{Introduction}

In the heliosphere, ejection of huge amount of solar plasma ($10^ {15}$-$10^ {16}$ g) is caused by CME.  As seen in coronagraph images, the three part structure of a CME (Alexander et al. 2006), is probably maintained at 1 AU and is believed to be recognized as plasma pileup, magnetic flux rope and ejected cold plasma respectively (Sharma et al. 2012). Further the relation between disappearing filaments and ICMEs was well established by Wilson \& Hildner 1984, 1986. In our study we identified 3 ICMEs at 1AU using their solar wind in-situ observations and also examined how different ICMEs associated with filaments lead to moderate or major geomagnetic activity on their arrival at the Earth.

\section{Observations and Analysis}

We have studied 3 filament eruption events observed during 2007 to 2012 and selected the events from the list by Yan Li et al. (2014). First, we discuss only one case of CME i.e. 28 September 2012 in detail and the similar analysis is carried out for other two cases as listed in Table \ref{tab1}.
 
\begin{table}
  \begin{center}
  \caption{From left to right: the appearance of CMEs in LASCO, their speeds, 1 AU arrival times, 1 AU magnetic field, induced geomagnetic storm index (Dst) and location of CMEs associated erupted filament.}
  \label{tab1}
 {\scriptsize
  \begin{tabular}{|l|c|c|c|c|c|c|c|}\hline 
{\bf CME } & {\bf Type of } & {\bf Linear  } & {\bf Arrival  } & {\bf ICME  } & {\bf B } & {\bf Dst } & {\bf Solar } \\ 
   {\bf Date }&  {\bf CME (halo/ } & {\bf speed } & {\bf of } & {\bf Start time } &{\bf (nT)} & {\bf (nT)} & {\bf location }\\ 
	{\bf and Time}&  {\bf Partial halo)} & {\bf (km/s)} & {\bf Disturbance} & {\bf End} & & & {\bf of Filament}\\ \hline
	
23 May  & halo & 258 & 28 May  & 28 May 2010 at  & 13.5 & -80 & N20E10 \\
   2010 at & & & 2010 at & 19:00 UT to 29 May & & &  \\ 
	18:06 UT & & & 02:58 UT & 2010 at 17:00 UT & & &  \\\hline
	
6 Oct & Partial  & 282 & 11 Oct  & 11 Oct 2010 at  & 14.3 & -70 & N30E23 \\
   2010 at & halo & & 2010 at & 05:50 UT to 11 Oct & & &  \\ 
	7:12 UT & & & 5:50 UT & 2010 at 17:16 UT & & &  \\\hline	
 
23 Sep  & halo & 947 & 30 Sep  & 1 Oct 2012 at  & 14 & -122 & N09W26 \\
   2012 at & & & 2012 at & 00:00 UT to 2 Oct & & &  \\ 
	00:12 UT & & & 23:05 UT & 2012 at 00:00 UT & & &  \\\hline

 \end{tabular}
  }
 \end{center}
 \end{table}

\begin{figure}[h]
	\begin{center}
		\includegraphics[scale=0.4]{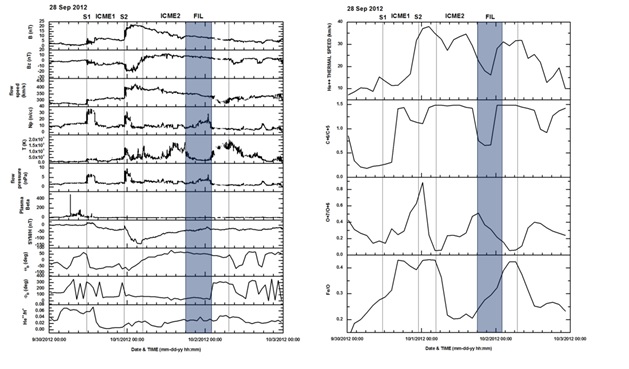}
		\caption{The label S1 and S2 mark the arrival of shocks associated with the CME1 and CME2, respectively. The shaded part is possible location of filament material associated with CME2. }
		\label{Fig1}
	\end{center}
\end{figure}

{\underline{\it 28 September 2012 CME }}. We found that CME of 28 Sep 2012 arrived at 1AU following the arrival of another CME of 25 Sep (termed as CME1) which was launched from the Sun with a speed of 335 km $s^ {- 1}$ at 9:37 UT. The CME of 28 Sep (termed as CME2), appeared as a halo CME was launched at around 00 UT having a plane-of-sky speed of 950 km $s^ {- 1}$. Identiﬁcation of CMEs and their signatures of interaction, from in situ data analysis, are in accordance with that reported by Liu et al. (2014) and Mishra et al. (2015). Here, we focus on the identification of filament material at 1 AU. The in-situ plasma and magnetic field plots show an enhancement in density up to 25 $cm^ {- 3}$ and depression in temperature up to 2 $\times$ $10 ^ {4}$K is observed at the rear edge of CME2. The $He{++}$ \slash $H{+}$ ratio increased and C+6 \slash C+5 ratio decreased up to 8\% and 0.6 respectively, thermal speed of helium ion and avg charge state of iron dropped to 20 km $s^ {- 1}$ and up to 9 respectively (figure \ref{Fig1}). Therefore, based on these signatures, we confirm the identiﬁcation of filament material at 1 AU at the rear edge of CME2 as reported by previous studies (Mishra et al., 2015).

\section{Results and Conclusion}
The filament material identified in all the three interplanetary CMEs at 1 AU are cold or low in temperature (of the order of $10 ^ {4}$ K) compared to the other structures in the ICME. During the region of filament, we find low charge states as observed by a decrease in charge states of carbon, oxygen and iron except in one case of 28 Sept. CMEs which do not show low iron charge states. $He{++}$ \slash $H{+}$ abundance ratio increase and decrease in thermal velocities was observed for filaments identified in 2 cases as expected while in the case of May 2010 there was a decrease in $He{++}$ \slash $H{+}$ abundance ratio \& increase in thermal velocities.  This is possibly because of interaction of two CMEs in this case.


\begin{thebibliography}{}

\bibitem[Amari \etal\ (1995)]{Amari_etal95}
{Li, Y., J. G. Luhmann, B. J. Lynch, \& E. K. J. Kilpua } 2014,
\textit{J. Geophys. Res. Space Physics}, 119, 3237  

\bibitem[Anders \& Zinner (1993)]{AndersZinner93}
{Mishra, W., N. Srivastava, \& T. Singh} 2015, 
\textit{J. Geophys. Res. Space Physics}, 120, 221


\end{thebibliography}
\end{document}